\documentclass[11pt,showkeys,aps,pra]{revtex4}
\usepackage[utf8x]{inputenc}
\usepackage{ucs}

\usepackage{amsmath,amssymb}
\usepackage{latexsym}
\usepackage{amsfonts}
\usepackage{dcolumn}
\usepackage{bm}
\usepackage[usenames]{color}
\usepackage{multirow}
\usepackage{graphicx}
\usepackage{hyperref}
\usepackage{subfig}
\usepackage{booktabs}
\usepackage{xcolor}
\usepackage[normalem]{ulem}
\usepackage{float} 
\usepackage{wrapfig} 
\usepackage{upgreek} 
\usepackage{cancel} 
\usepackage{mathdots} 
\usepackage{mathrsfs} 
\graphicspath{{figures/}} 

\renewcommand{\arraystretch}{1}

\begin{document}

\title{R\'enyi entropies for multidimensional hydrogenic systems in position and momentum spaces}

\author{D. Puertas-Centeno}
\email[]{vidda@correo.ugr.es}
\affiliation{Departamento de F\'{\i}sica At\'{o}mica, Molecular y Nuclear, Universidad de Granada, Granada 18071, Spain}
\affiliation{Instituto Carlos I de F\'{\i}sica Te\'orica y Computacional, Universidad de Granada, Granada 18071, Spain}

\author{I.V. Toranzo}
\email[]{ivtoranzo@ugr.es}
\affiliation{Departamento de F\'{\i}sica At\'{o}mica, Molecular y Nuclear, Universidad de Granada, Granada 18071, Spain}
\affiliation{Instituto Carlos I de F\'{\i}sica Te\'orica y Computacional, Universidad de Granada, Granada 18071, Spain}

\author{J.S. Dehesa}
\email[]{dehesa@ugr.es}
\affiliation{Departamento de F\'{\i}sica At\'{o}mica, Molecular y Nuclear, Universidad de Granada, Granada 18071, Spain}
\affiliation{Instituto Carlos I de F\'{\i}sica Te\'orica y Computacional, Universidad de Granada, Granada 18071, Spain}

\begin{abstract}
The R\'enyi entropies of Coulomb systems $R_{p}[\rho], 0 < p < \infty$ are logarithms of power functionals of the electron density $\rho(\vec{r})$ which quantify most appropriately the electron uncertainty and describe numerous physical observables. However, its analytical determination is a hard issue not yet solved except for the first lowest-lying energetic states of some specific systems. This is so even for the $D$-dimensional hydrogenic system, which is the main prototype of the multidimensional Coulomb many-body systems. Recently, the R\'enyi entropies of this system have been found in the two extreme high-energy (Rydberg) and high-dimensional (pseudo-classical) cases. In this work we determine the position and momentum R\'enyi entropies (with integer $p$ greater than 1) for all the discrete stationary states of the multidimensional hydrogenic system directly in terms of the hyperquantum numbers which characterize the states, the nuclear charge and the space dimensionality. We have used a methodology based on linearization formulas for powers of the orthogonal Laguerre and Gegenbauer polynomials which control the hydrogenic states.
\end{abstract}


\maketitle

\section{Introduction}

In a seminal paper Alfr\'ed R\'enyi \cite{renyi1} found axiomatically a set of monoparametric information entropies of a probability density $\rho(\vec{x})$ which includes the Shannon entropy as a limiting case. These R\'enyi quantities are logarithms of integral functionals of powers of $\rho(\vec{x})$ (Yule-Sichel frequency moments \cite{yule,sichel1,sichel2}) appropriately renormalized to have an entropic character, as
\begin{equation}
\label{Renyidef}
R_{q}[\rho] := \frac{1}{1-q}\ln \int_{\mathbb{R}^{D}} [\rho(\vec{x})]^{q}\, d\vec{x} , \quad \vec{x}\in \mathbb{R}^{D},\, q\neq 1
\end{equation}
These entropies, which completely characterize the density under certain conditions, quantify various spreading-like facets (governed by the parameter $q$) of the probability density $\rho(\vec{x})$, including the Shannon entropy (when $q\rightarrow 1$) and the disequilibrium (when $q=2$) which measures the separation of the distribution with respect to equiprobability. Moreover, when $q\rightarrow 0$ this quantity is proportional to the logarithm of the volume of the multidimensional support set, and when $q\rightarrow \infty$ the Rényi entropy puts more emphasis on where $\rho(\vec{x})$ attains its maximum. The parameter $q$ has different meanings depending on the context; for instance, it can be interpreted as the inverse of the temperature in thermodynamic systems and it is related to the Reynolds number in turbulence theory \cite{sen2012}.
Moreover, the R\'enyi entropies are closely related to other information-theoretic quantities such as e.g., the Tsallis entropies \cite{tsallis} which play a very important role in systems with strong long-range correlations and nonextensive statistical mechanics \cite{tsallis1,abe}. Furthermore, from the viewpoint of thermodynamics, the Rényi entropy its directly connected to the free energy of a system in thermal equilibrium, a relation that holds for both classical and quantum realms, and so accounts for the work that the system is capable of.   \\

The properties of the R\'enyi entropies and their applications have been widely considered/applied (see e.g., \cite{aczel,jizba2004a,jizba_2004b,rosso,brody,leonenko,nielsen,aptekarev2012a,aptekarev2012b,sanchez2013,calixto,carrillo,dehesa2015,aptekarev2016}) and reviewed \cite{dehesa_sen12,bialynicki3,jizba,portesi2018} in a broad variety of fields ranging from applied mathematics, quantum physics, Rydberg physics, complexity theory to non-linear physics, option price calibration, nanotechnology and neuroscience. However, these quantities have not yet been exactly calculated except for a few one-dimensional exponential densities (see e.g., \cite{nielsen}) and some probability densities of a single-particle system moving in the elementary multidimensional quantum potentials of infinite well \cite{aptekarev2012a} and rigid rotator \cite{dehesa2015}) types.  Moreover, the dominant term for the R\'enyi entropies of the multidimensional harmonic oscillator has been determined  at the high-dimensional (pseudoclassical) and high-energy (Rydberg) limits \cite{dehesa2017}, and then the entropy values for both ground and excited oscillator-like states have been analytically calculated \cite{puertas2018} in terms of the hyperquantum numbers and the oscillator strength. Here we should also mention the efforts to understand the role of quantum entanglement of many-body systems by determining the quantum R\'enyi entropy of one-dimensional many-particle model systems under special and limiting conditions, such as finite systems of free fermions in a continuum interval, statistical models with finite numbers of degrees of freedom, the one-dimensional XY quantum spin chain in a transverse magnetic field and the totally asymmetric exclusion process \cite{francini2008, calabrese2009, calabrese2011, wood2017}, among others.\\

Recently, the analytical determination of the R\'enyi entropies of the main prototype of the $D$-dimensional Coulomb many-body systems, the $D$-dimensional hydrogenic system, from first principles (i.e., in terms of the hyperquantum numbers of the state and the nuclear charge) has been undertaken \cite{dehesa2010,tor2017b,tor2016b}. This is relevant \textit{per se} and for a reference point of view for all multidimensional Coulomb systems. The $D$-dimensional hydrogenic system is a negatively-charged particle moving in a space of $D$ dimensions around a positively charged core which electromagnetically binds it in its orbit \cite{nieto,yanez1994,kostelecky,dehesa2010,aquilanti,coletti,caruso,bures}. This system allows for the modelling of numerous three-dimensional physical systems (e.g., hydrogenic atoms and ions, exotic atoms, antimatter atoms, Rydberg atoms) and a number of nanotechnological objects (quantum wells, wires and dots) and qubits  which have been shown to be very useful in semiconductor physics \cite{harrison, li:pla07} and quantum technologies \cite{nieto:pra00, dykman:prb03}, respectively. Moreover, it plays a crucial role for the interpretation of numerous phenomena of quantum cosmology \cite{amelinocamelia_05} and quantum field theory \cite{witten, itzykson_06,dong}. In addition, the $D$-dimensional hydrogenic wavefunctions have been used as complete orthonormal sets for many-body atomic and molecular problems \cite{aquilanti:aqc01,coletti} in both position and momentum spaces. However, although tremendous advances have been witnessed in understanding the (energetic) spectroscopic properties of three-dimensional hydrogenic atoms, the informational properties of standard and non-standard hydrogenic systems are barely known. The latter properties, which crucially depend on the system's eigenfunctions, quantify the various facets of the spatial extension or multidimensional spreading of the electronic charge. The aim of this work is to cover this informational lack by means of the determination of the R\'enyi entropies for the multidimensional hydrogenic system.\\

The calculation of the hydrogenic R\'enyi entropies is a difficult task except for the lowest-lying energy states. This is because these quantities are described by means of some power or logarithmic functionals of the electron density, which cannot be easily handled in an analytical way nor numerically computed; the latter is basically because a naive numerical evaluation using quadratures is not convenient due to the increasing number of integrable singularities when the principal hyperquantum number $n$ is increasing, which spoils any attempt to achieve reasonable accuracy even for rather small $n$. Up until now, these quantities have been only calculated in a compact form \cite{tor2017b,toranzo16a,tor2016b} at the high-dimensional (pseudoclassical) and high-energy (Rydberg) limits by use of modern asymptotical techniques of the Laguerre and Gegenbauer  polynomials which control the state's wavefunctions in position and momentum spaces \cite{aptekarev2016,temme2017}.\\

In this work we determine the R\'enyi entropies $R_{q}[\rho]$ (with integer $q$ greater than 1) for the electron density $\rho(\vec{r})$ of all the discrete stationary states of the $D$-dimensional hydrogenic system directly in terms of the hyperquantum numbers which characterize the states, the nuclear charge and the space dimensionality $D$. The structure of the manuscript is the following. In Sec. 2 the notion of the $q$th-order R\'enyi entropy for a $D$-dimensional probability is given, and then the wavefunctions of the hydrogenic states in the $D$-dimensional configuration space are briefly described so as to express the associated probability densities. In Sec. 3 the position and momentum R\'enyi entropies are analytically determined by means of the little known polynomial linearization methodology of Srivastava-Niukkanen type \cite{srivastava,niukkanen,niukkanen2}. In Sec. 4 the specific values for the entropies of some particularly relevant hydrogenic states are given to illustrate the applicability of our procedure. Finally, some concluding remarks and open problems are given.

\section{$D$-dimensional hydrogenic system: An entropic view}

In this section we briefly describe the quantum position and momentum probability setting of the $D$-dimensional hydrogenic system where the R\'enyi entropies are applied. For convenience we start with the definition of these entropies for a general multidimensional probability density, and then we give the known wavefunctions \cite{nieto,yanez1994,kostelecky,aquilanti} of the system in both position and momentum spaces as well as the corresponding quantum probability densities.

\subsection{Rényi entropy}

The Rényi entropies $R_{q}[\rho]$  of a $D$-dimensional probability density $\rho(\vec{r})$ are defined as
\begin{eqnarray}
\label{eq:renentrop}
R_{q}[\rho] &=&  \frac{1}{1-q}\ln W_{q}[\rho]; \quad 0<q<\infty,\quad q \neq 1,
\end{eqnarray}
where $W_{q}[\rho]$ denotes the entropic or Yule-Sichel frequency moment of order $q$ of $\rho(\vec{r})$ is given by 
\begin{equation}
\label{eq:entropmom}
W_{q}[\rho] = \int_{\mathbb{R}^D} [\rho(\vec{r})]^{q}\, d\vec{r} =\| \rho\|_q^q;\quad q > 0, 
\end{equation}
where the  position $\vec{r}=(x_1 ,  \ldots  , x_D)$ is  given in hyperspherical units as $(r,\theta_1,\theta_2,\ldots,\theta_{D-1}) \equiv (r,\Omega_{D-1})$, $\Omega_{D-1}\in S^{D-1}$; and the volume element is 
\begin{equation}
\label{dvol}
d\vec{r} =r^{D-1}drd\Omega_{D} , \quad d\Omega_{D-1} = \left(\prod_{j=1}^{D-2}\sin^{2\alpha_{j}}\theta_{j}\,d\theta_{j}\right)d\phi,
\end{equation}
with $2\alpha_{j}= D-j-1$. We have used $r \equiv |\vec{r}| = \sqrt{\sum_{i=1}^D x_i^2}
\in [0  \: ;  \: +\infty)$  and $x_i =  r \left(\prod_{k=1}^{i-1}  \sin \theta_k
\right) \cos \theta_i$ for $1 \le i \le D$
and with $\theta_i \in [0 \: ; \: \pi), i < D-1$, $\theta_{D-1} \equiv \phi \in [0 \: ; \: 2
\pi)$. By  convention $\theta_D =  0$ and the  empty product is the  unity. 

\subsection{Hydrogenic system}

The discrete stationary states of the $D$-dimensional hydrogenic system (i.e., a particle moving in the Coulomb potential $V_{D}(r) = - \frac{Z}{r}$, where $Z$ denotes the nuclear charge; atomic units are used throughout the paper) are known to be expressed \cite{yanez1994,dehesa2010} in position space by the energy eigenvalues
\begin{equation} \label{eqI_cap1:energia}
E= -\frac{Z^2}{2\eta^2},\hspace{0.5cm} \hspace{0.5cm} \eta=n+\frac{D-3}{2}; \hspace{5mm} n=1,2,3,...,
\end{equation}
and the associated eigenfunctions
\begin{eqnarray}
\label{eq:wavpos}
\Psi_{n,l,\{\mu\}}(\vec{r}) &=&
N_{n,l}\left(\frac{r}{\lambda}\right)^{l}e^{-\frac{r}{2\lambda}}\,\mathcal{L}_{n-l-1}^{(2l+D-2)}\left(\frac{r}{\lambda}\right)\, \mathcal{Y}_{l,\{\mu\}}(\Omega_{D-1})\nonumber\\
&=& N_{\eta,l}\left[\frac{\omega_{2L+1}(\tilde{r})}{\tilde{r}^{D-2}}\right]^{1/2}\mathcal{L}_{\eta-L-1}^{(2L+1)}(\tilde{r})\, \mathcal{Y}_{l,\{\mu\}}(\Omega_{D-1}),
\end{eqnarray}
with 
\begin{eqnarray}
\eta &=& n + \frac{D-3}{2}, \quad n=1,2,3,\ldots\nonumber \\
\label{eq:1}
L &=& l+\frac{D-3}{2}, \quad l = 0,1,2,\ldots\nonumber \\
\label{eq:2}
\tilde{r} &=& \frac{r}{\lambda}\quad with\quad \lambda=\frac{\eta}{2Z},
\label{eq:3}
\end{eqnarray}
The symbol $\eta$ denotes the principal hyperquantum number of the state associated to the radial coordinate, and $(l,\left\lbrace \mu \right\rbrace)\equiv(l\equiv\mu_1,\mu_2,...,\mu_{D-1})$ denote the orbital and magnetic hyperquantum numbers associated to the angular variables $\Omega_{D-1}\equiv (\theta_1, \theta_2,...,\theta_{D-1})$, which may take all values consistent with the inequalities $l\equiv\mu_1\geq\mu_2\geq...\geq \left|\mu_{D-1} \right| \equiv \left|m\right|\geq 0$.
In addition,
$\omega_{\alpha}(x) =x^{\alpha}e^{-x}, \, \alpha=2l+D-2$
  is the weight function of the orthogonal and orthonormal Laguerre polynomials \cite{nikiforov,olver} of degree $n$ and parameter $\alpha$, here denoted by $L_{n}^{(\alpha)}(x)$ and $\widehat{L}_{n}^{(\alpha)}(x)$, respectively and
\begin{equation}
\label{eq:4}
N_{n,l} = \lambda^{-\frac{D}{2}}\left(\frac{(\eta-L-1)!}{2\eta(\eta+L)!}\right)^{\frac{1}{2}}
\end{equation}
is the normalization constant which ensures the unit norm of the wavefunction.
The angular part of the eigenfunctions is given by the hyperspherical harmonics as
\begin{equation}
\mathcal{Y}_{l,\{\mu\}}(\Omega_{D-1}) = \mathcal{N}_{l,\{\mu\}}e^{im\phi}\times \prod_{j=1}^{D-2}\mathcal{C}^{(\alpha_{j}+\mu_{j+1})}_{\mu_{j}-\mu_{j+1}}(\cos\theta_{j})(\sin\theta_{j})^{\mu_{j+1}}
\label{eq:hyperspherarm}
\end{equation}
where $\mathcal{N}_{l,\{\mu\}}$ is the normalization constant
\begin{equation}
\label{eq:normhypersphar}
\mathcal{N}_{l,\{\mu\}}^{2} = \frac{1}{2\pi}\times
\prod_{j=1}^{D-2} \frac{(\alpha_{j}+\mu_{j})(\mu_{j}-\mu_{j+1})![\Gamma(\alpha_{j}+\mu_{j+1})]^{2}}{\pi \, 2^{1-2\alpha_{j}-2\mu_{j+1}}\Gamma(2\alpha_{j}+\mu_{j}+\mu_{j+1})},\nonumber\\
\end{equation}
the symbol $\mathcal{C}^{(\lambda)}_{n}(t)$ denotes the Gegenbauer polynomial \cite{nikiforov,olver} of degree $n$ and parameter $\lambda$, and $2\alpha_j=D-j-1$.\\

Then, the position probability density of a $D$-dimensional hydrogenic state characterized by the hyperquantum numbers $(n,l,\{\mu\})$ is given by the squared modulus of the position eigenfunction as 
\begin{eqnarray}
\label{eq:denspos}
\rho_{n,l,\{\mu\}}(\vec{r}) &=& 
N_{\eta,l}^{2}\left[\frac{\omega_{2L+1}(\tilde{r})}{\tilde{r}^{D-2}}\right][\mathcal{L}_{\eta-L-1}^{(2L+1)}(\tilde{r})]^{2}\, |\mathcal{Y}_{l,\{\mu\}}(\Omega_{D-1})|^{2}
 \nonumber \\
&=& 
N_{n,l}^{2}\tilde{r}^{2l}e^{-\tilde{r}}[\mathcal{L}_{n-l-1}^{(2l+D-2)}(\tilde{r})]^{2}\, |\mathcal{Y}_{l,\{\mu\}}(\Omega_{D-1})|^{2}
\nonumber \\
&\equiv& \rho_{n,l}(\tilde{r})\,|\mathcal{Y}_{l,\{\mu\}}(\Omega_{D-1})|^{2}.
\end{eqnarray}

Moreover, the Fourier transform of the position eigenfunction $\Psi_{\eta,l, \left\lbrace \mu \right\rbrace }(\vec{r})$ given by (\ref{eq:wavpos}), i.e., $\tilde{\Psi}(\vec{p})=\int_{\mathbb{R}^{D}}e^{-i\vec{p}\cdot\vec{r}}\Psi(\vec{r})\frac{d\vec{r}}{(2\pi)^{\frac{D}{2}}}$, provides the eigenfunction of the system in the momentum space as
\begin{equation}
\label{eq:momwvf}
\tilde{\Psi}_{n,l,\{\mu\}}(\vec{p}) = \mathcal{M}_{n,l}(p)\,\,\mathcal{Y}_{l,\{\mu \}}(\Omega_{D-1}),
\end{equation}
where the radial part is 
\begin{equation}
\label{eq:radmomwvf}
\mathcal{M}_{n,l}(p) = K_{n,l}\frac{(\eta \tilde{p})^{l}}{(1+\eta^{2}\tilde{p}^{2})^{L+2}}\,\mathcal{C}_{\eta-L-1}^{(L+1)}\left( \frac{1-\eta^{2}\tilde{p}^{2}}{1+\eta^{2}\tilde{p}^{2}} \right)
\end{equation}
with $\tilde{p} =\frac{p}{Z}$ and the normalization constant
\begin{equation}
\label{eq:normacons}
K_{n,l} = Z^{-\frac{D}{2}}2^{2L+3}\left[\frac{(\eta-L-1)!}{2\pi(\eta+L)!} \right]^{\frac{1}{2}}\Gamma(L+1)\eta^{\frac{D+1}{2}}.
\end{equation}
Then, the momentum probability density of the $D$-dimensional hydrogenic stationary state with the hyperquantum numbers $(n,l,\{\mu\})$ is 
\begin{eqnarray}
\label{eq:momdens}
\gamma_{n,l,\{\mu\}}(\vec{p}) &=& |\tilde{\Psi}_{n,l,\{\mu \}}(\vec{p})|^{2} =  \mathcal{M}^{2}_{n,l}(p) \,\,|\mathcal{Y}_{l,\{\mu \}}(\Omega_{D-1})|^{2}\nonumber \\
&&
\hspace{-2cm}= K_{n,l}^{2}\frac{(\eta \tilde{p})^{2l}}{(1+\eta^{2}\tilde{p}^{2})^{2L+4}}\left[\mathcal{C}_{\eta-L-1}^{(L+1)}\left( \frac{1-\eta^{2}\tilde{p}^{2}}{1+\eta^{2}\tilde{p}^{2}} \right)\right]^{2}|\mathcal{Y}_{l,\{\mu \}}(\Omega_{D-1})|^{2}.
\end{eqnarray}
Note that the position and momentum probability densities are quite different since the radial parts of the densities are controlled by the Laguerre and the Gegenbauer polynomials in the position and momentum cases, respectively. Then, we expect the R\'enyi entropies of the hydrogenic system in the position and momentum spaces to be different; this is not surprising because the R\'enyi entropies are not observables of the system. They  satisfy the known R\'enyi-entropy-based uncertainty relation \cite{jizba,bialynicki2,vignat,portesi} as given by Eq. (\ref{sum}) below. The latter relation indicates that the joint position-momentum R\'enyi entropy, which is given by the sum of the position and momentum R\'enyi entropies, is bounded from below for all bound stationary states of the system.

\section{Exact Rényi entropies of the hydrogenic system}
In this section we determine the position and momentum R\'enyi entropies  $R_{q}[\rho]$ (with natural $q$ other than unity) for all the discrete stationary states of the $D$-dimensional hydrogenic system in an analytical way. First we note that these entropies can be decomposed into two radial and angular parts in both conjugated spaces. Then, we use a recent procedure \cite{sanchez2013} based on the Srivastava-Niukkanen method \cite{srivastava,niukkanen,srivastava2} which linearize integer powers of Laguerre and Jacobi polynomials. The involved linearization coefficients are expressed via some multiparametric hypergeometric functions of Lauricella and Srivastava-Daoust types, respectively.\\

From Eqs. (\ref{eq:renentrop}), (\ref{eq:entropmom}) and (\ref{eq:denspos}) the Rényi entropies of the $D$-dimensional hydrogenic state $(n,l,\{\mu\})$ in position space can be written as
\begin{equation}
\label{eq:renyihyd1}
R_{q}[\rho_{n,l,\{\mu\}}] = R_{q}[\rho_{n,l}]+R_{q}[\mathcal{Y}_{l,\{\mu\}}],
\end{equation}
where $R_{q}[\rho_{n,l}]$ denotes the radial part
\begin{equation}
\label{eq:renyihyd2}
R_{q}[\rho_{n,l}] = \frac{1}{1-q}\ln \int_{0}^{\infty} [\rho_{n,l}]^{q} r^{D-1}\, dr,
\end{equation}
and  $R_{q}[\mathcal{Y}_{l,\{\mu\}}]$ denotes the angular part
\begin{equation}
\label{eq:renyihyd3}
R_{q}[\mathcal{Y}_{l,\{\mu\}}] = \frac{1}{1-q}\ln \Lambda_{l,\{\mu\}}(q),
\end{equation}
with
\begin{equation}
\label{angpart}
\Lambda_{l,\{\mu\}}(q) = \int |\mathcal{Y}_{l,\{\mu\}}(\Omega_{D-1})|^{2q}\, d\Omega_{D-1}.
\end{equation}

\subsection{Radial R\'enyi entropy in position space}

Taking into account Eqs. (\ref{eq:denspos}) and (\ref{eq:renyihyd2}), the radial Rényi entropy can be written as
{\small
\begin{eqnarray}
\label{eq:rrd1}
R_{q}[\rho_{n,l}] &=& 
\frac{1}{1-q}\ln \left[ N_{n,l}^{2q} \int_{0}^{\infty} \tilde{r}^{2lq}e^{-q\tilde{r}}[\mathcal{L}_{n-l-1}^{(2l+D-2)}(\tilde{r})]^{2q}\, r^{D-1}dr\right], \nonumber  
\\
&=& \frac{1}{1-q}\ln \left[ \lambda^{D(1-q)}\left(\frac{\Gamma(n-l)}{2\eta\Gamma(n+l+D-2)}\right)^{q} \right] \nonumber\\
& &\hspace{-0.5cm} + \frac{1}{1-q}\ln \int_{0}^{\infty} \tilde{r}^{2lq+D-1}e^{-q\tilde{r}}[\mathcal{L}_{n-l-1}^{(2l+D-2)}(\tilde{r})]^{2q}\,d\tilde{r}. 
\end{eqnarray}
}
To evaluate the integral first we perform the change of variable $x=q\tilde{r}$ to have
\begin{eqnarray}
\label{eq:rrd2}
R_{q}[\rho_{n,l}] &=& \frac{1}{1-q}\ln \left[ \lambda^{D(1-q)}\left(\frac{\Gamma(n-l)}{2\eta\Gamma(n+l+D-2)}\right)^{q} \right] \nonumber \\
& &\hspace{-2cm}+ \frac{1}{1-q}\ln q^{-D-2lq}\int_{0}^{\infty} x^{2lq+D-1}e^{-x}\left[\mathcal{L}_{n-l-1}^{(2l+D-2)}\left(\frac{x}{q}\right)\right]^{2q}\,dx,
\end{eqnarray}
and then we apply the linearization formula of the $(2q)$th-power of the Laguerre polynomial $L_{n-l-1}^{(2l+D-2)}\left(\frac{x}{q}\right)$ given by
\begin{equation}
 \label{linlag}
      y^a \left[\mathcal{L}_{k}^{(\alpha)}\left(ty\right)\right]^{r} = \sum_{i=0}^{\infty} c_{i}\left(a,r,t,k,\alpha,\gamma\right) \mathcal{L}_{i}^{(\gamma)}(y),
 \end{equation}
with $a>0, t>0, \alpha > -1, \gamma > -1$, the integer $k \geq0, i\geq 0$, and the linearization coefficients 
 \begin{eqnarray}
  \label{linlagcoeff1}
  c_{i}\left(a,r,t,k,\alpha,\gamma\right) =(\gamma+1)_a  \left(\frac{\Gamma(k+\alpha+1)}{\Gamma(\alpha+1)\Gamma(k+1)}\right)^r & & \nonumber \\
&   & \hspace{-7cm}\times F_{A}^{(r+1)}\left( \begin{array}{cc}
  							\gamma+a+1; \overbrace{-k, \ldots, -k}^{r},-i & \\
  																			&; \underbrace{t,\ldots,t}_r,1\\
  							\underbrace{\alpha+1, \ldots, 	\alpha+1}_r,\gamma+ 1 & \\
  							\end{array}\right),
   \end{eqnarray}
where the Pochhammer symbol $(z)_a = \frac{\Gamma(z+a)}{\Gamma(z)}$ and the symbol $F_{A}^{(s)}(x_{1},\ldots,x_{r})$ denotes the Lauricella function of type A of $s$ variables and $2s+1$ parameters defined as \cite{srivastava2}
 \begin{equation}
 \label{laurifunc}
 F_{A}^{(s)}\left( \begin{array}{cc}
  							a; b_{1}, \ldots, b_{s} & \\
  																			&; x_{1},\ldots, x_{s}\\
  							c_{1}, \ldots, 	c_{s} & \\
  							\end{array}\right) =\sum_{j_{1},\ldots, j_{s}=0}^{\infty} \frac{(a)_{j_{1}+\ldots+j_{s}}(b_{1})_{j_{1}} \cdots (b_{s})_{j_{s}}}{(c_{1})_{j_{1}} \cdots (c_{s})_{j_{s}}} \frac{x_{1}^{j_{1}}\cdots x_{s}^{j_{s}} }{j_{1}!\cdots j_{s}! }.
 \end{equation}
  
  Now, taking $a=2lq+D-1$, $r=2q$, $t=\frac1q$, $k=n-l-1$, $\alpha= 2l+D-2$, inserting (\ref{linlag}) in the integral kernel of (\ref{eq:rrd2}) and using the orthogonalization condition of the Laguerre polynomials \cite{olver}, after some algebraic manipulations one finds that the final expression of the radial Rényi entropy is given by
\begin{eqnarray}\label{HSRRE2}
\hspace{-3cm}R_{q}[\rho_{n,l}]=D\ln \left(\frac{\eta}{2Z}\right)+\frac{q}{1-q}\ln\left(\frac{(\eta-L)_{2L+1}}{2\eta}  \right) + \frac{1}{1-q}\ln\mathcal F_q(D,\eta,L)  +\frac{1}{1-q}\ln\mathcal A_q(D,L)\,\quad\quad
\end{eqnarray}
where 
\begin{eqnarray}
\label{Fdef}
\mathcal F_q(D,n,l)  &  &\equiv F_{A}^{(2q)}\left( \begin{array}{cc}
  							2lq+D; \overbrace{-n+l+1, \ldots, -n+l+1}^{2q} & \\
  																			&; \underbrace{\frac1q,\ldots,\frac1q}_{2q}\\[-1.5em]
  							\underbrace{2l+D-1, \ldots, 	2l+D-1}_{2q} & \\
  							\end{array}\right),\quad\quad\quad
\end{eqnarray}
and $\mathcal A_q(D,L)\equiv\frac{\Gamma\left(D+2lq\right)}{q^{D+2lq}\Gamma\left(2L+2\right)^{2q}}.  $
Note that when $l=n-1$ the function $\mathcal F_q(D,n,l)$ is equal to unity,  so that the third term of the entropy expression (\ref{HSRRE2}) vanishes. Moreover, let us highlight that, from Eq. (\ref{laurifunc}), this function defines a finite sum by taking into account the properties of the involved Pochhammer symbols with negative integer arguments.
\subsection{Angular Rényi entropy}

Now, from Eqs. (\ref{eq:renentrop}), (\ref{eq:entropmom}), (\ref{eq:renyihyd3}), (\ref{angpart}) and (\ref{eq:hyperspherarm}) one has that the angular Rényi entropy has the form
\begin{equation}
\label{angren1}
R_{q}[\mathcal{Y}_{l,\{\mu\}}] = \frac{1}{1-q}\ln \int \mathcal{N}_{l,\{\mu\}}^{2q}  \prod_{j=1}^{D-2}[\mathcal{C}^{(\alpha_{j}+\mu_{j+1})}_{\mu_{j}-\mu_{j+1}}(\cos\theta_{j})]^{2q}|\sin\theta_{j}|^{2q\mu_{j+1}}\,d\Omega_{D-1}
\end{equation}
With the change of variable $t=\cos\theta_{j}$, this integral can be rewritten as
\begin{equation}
\label{angren2}
R_{q}[\mathcal{Y}_{l,\{\mu\}}] 
=\frac{1}{1-q}\ln \left(2\pi\mathcal{N}_{l,\{\mu\}}^{2q} \right) + \frac{1}{1-q}\ln \left[\prod_{j=1}^{D-2} \mathcal{I}_{j}(q)  \right],
\end{equation}
where 
\begin{equation}
\label{gegint}
\mathcal I_q(D,\mu_j,\mu_{j+1})=\int_{-1}^{1} [\mathcal{C}^{(\alpha_{j}+\mu_{j+1})}_{\mu_{j}-\mu_{j+1}}(t)]^{2q}|(1-t^{2})|^{q\mu_{j+1}+\alpha_{j}-\frac{1}{2}}\,dt.
\end{equation}
To calculate this integral we use the known relationship between the Gegenbauer and Jacobi polynomials \cite{olver},
\begin{equation}
\label{relgegjac}
\mathcal{C}_{\kappa}^{(\lambda)}(x) = \frac{\Gamma\left(\lambda+\frac{1}{2}\right)}{\Gamma(2\lambda)}\frac{\Gamma(\kappa+2\lambda)}{\Gamma\left(\kappa+\lambda+\frac{1}{2}\right)} \mathcal{P}_{\kappa}^{\left(\lambda-\frac{1}{2},\lambda-\frac{1}{2}\right)}(x)
\end{equation} 
together with the Srivastava-Niukkanen-based linearization formula of the Jacobi polynomials \cite{sanchez2013}
\begin{equation}
\label{linjac}
[\mathcal{P}_{\kappa}^{(\alpha,\beta)}(x)]^{2q} = \sum_{i=0}^{\infty} \tilde c_{i}(0,2q,\kappa,\alpha,\beta,\gamma,\delta) \mathcal{P}_{i}^{(\gamma,\delta)}(x),
\end{equation}
with $\alpha>-1, \beta>-1, \gamma>-1, \delta>-1$ and where the linearization coefficients $\tilde c_i$ are given by
\begin{eqnarray}
\label{sumcoeffs}
\tilde c_{i}(0,2q,\kappa,\alpha,\beta,\gamma,\delta) &=& \left(\frac{\Gamma(\kappa+\alpha+1)}{\Gamma(\alpha+1)\,\Gamma(\kappa+1)}\right)^{2q}
\frac{\gamma+\delta+2i+1}{\gamma+\delta+i+1}\nonumber \\
&&\hspace{-4cm}\times \sum_{j_{1}, \ldots, j_{2q}=0 }^{\kappa}\sum_{j_{2q+1}=0}^{i} \frac{(\gamma+1)_{j_{1}+\ldots+j_{2q}+j_{2q+1}} }{(\gamma+\delta+i+2)_{j_{1}+\ldots+j_{2q}} }\nonumber \\
&&\hspace{-4cm} \times \frac{(-\kappa)_{j_{1}}(\alpha+\beta+\kappa+1)_{j_{1}}\cdots (-\kappa)_{j_{2q}}(\alpha+\beta+\kappa+1)_{j_{2q}}(-i)_{j_{2q+1}} }{(\alpha+1)_{j_{1}}\cdots (\alpha+1)_{j_{2q}}(\gamma+1)_{j_{2q+1}}j_{1}!\cdots j_{2q}!j_{2q+1}! }.
\end{eqnarray}

Then, the orthogonalization relation of the Jacobi polynomials \cite{olver}
\begin{equation*}
\hspace{-3cm}\int_{-1}^1 (1-x)^\alpha (1+x)^\beta P_m^{(\alpha,\beta)}(x) P_n^{(\alpha,\beta)}(x)\,dx=\frac{2^{\alpha+\beta+1}}{n!} \frac{\Gamma(\alpha+n+1)\Gamma(\beta+n+1)}{(\alpha+\beta+2n+1)\Gamma(\alpha+\beta+n+1)}\,\delta_{m,n},
\end{equation*}
(for $\alpha,\beta>-1,$ and where $\delta_{n,m}$ denotes the Kronecker's delta function) reduces the infinite number of infinite terms of the sum involved in our Gegenbauer linearization to a single one: that for $i=0$. Then, after some algebraic manipulations one obtains that the analytical expression of the angular Rényi entropy is given by
\begin{eqnarray}\nonumber
\label{HSARE5}
\hspace{-1cm}R_{q}[\mathcal{Y}_{l,\{\mu \}}]  &=&\ln(2\pi^{\frac D2})+\frac{1}{1-q}\ln\left[\frac{\Gamma(l+\frac D2)^q}{\Gamma\left(ql+\frac D2\right)}\frac{\Gamma\left(qm+1\right)}{\Gamma(m+1)^q}\right]
\\
& & \hspace{-1cm}+\frac{1}{1-q} \sum_{j=1}^{D-2} \ln \left[\mathcal B_q(D,\mu_j,\mu_{j+1})\,\mathcal G_q\left(D,\mu_j,\mu_{j+1}\right)\right]
\end{eqnarray}
where 
\begin{equation}
\hspace{-3cm}\mathcal B_q\left(D,\mu_j,\mu_{j+1}\right)=\frac1{[(\mu_{j}-\mu_{j+1})!]^q}
\frac{(2\alpha_{j}+2\mu_{j+1}+1)_{2(\mu_j-\mu_{j+1})}^q}{(2\alpha_{j}+\mu_{j}+\mu_{j+1})_{\mu_j-\mu_{j+1}}^q} \frac{(q\mu_{j+1}+\alpha_{j}+1)_{q(\mu_j-\mu_{j+1})}}{(\alpha_{j}+\mu_{j+1}+1)_{\mu_j-\mu_{j+1}}^q} 
\end{equation}
and 
\begin{eqnarray}\nonumber
\label{tildec0}
\hspace{-1cm}\mathcal G_q(D,\mu_j,\mu_{j+1})&=& F_{1:1;\ldots;1}^{1:2;\ldots;2}\left( \begin{array}{cc}
				a_{j}: b_{j}, c_{j} ;\ldots;  b_{j}, c_{j} & \\
																&; 1, \ldots, 1\\
				d_{j}: e_{j}; \ldots; e_{j} & \\
				\end{array}\right)
\\\nonumber	
&=&\sum_{i_{1},\ldots,i_{2q}=0}^{\mu_{j}-\mu_{j+1}} \frac{(a_j)_{i_{1}+\ldots i_{2q} } }{(d_j)_{i_{1}+\ldots + i_{2q} } }\frac{ (b_j)_{i_{1} }(c_j)_{i_{1}} \cdots (b_j)_{i_{2q} }(c_j)_{i_{2q}}  }{(e_j)_{i_{1}}\cdots (e_j )_{i_{2q}} i_{1}!\cdots i_{2q}! }\nonumber\\
\end{eqnarray}
with  $a_j=\alpha_j+q\mu_{j+1}+\frac12$, $b_j=-\mu_j+\mu_{j+1}$, $c_j=2\alpha_j+\mu_{j+1}+\mu_j$, $d_j=2q\mu_{j+1}+2\alpha_j+1$ and  $e_j=\alpha_j+\mu_{j+1}+\frac12$.  Note that the sum becomes finite because $b_j$ is a negative integer number, and so $(b_j)_i=\frac{\Gamma(b_j+i)}{\Gamma(b_j)}=0, \quad \forall i>|b_j|$. 
Let us also highlight that when $\mu_j=\mu_{j+1}$, the function $\mathcal B_q(D,\mu_j,\mu_{j+1})=\mathcal G_q(D,\mu_j,\mu_{j+1})=1$. 
The symbol $F_{1:1;\ldots;1}^{1:2;\ldots;2}(x_1,\ldots,x_r)$ denotes the $r$-variate Srivastava--Daoust function \cite{srivastava,sanchez2013} defined as
\begin{eqnarray}
\label{daoust}
F_{1:1;\ldots;1}^{1:2;\ldots;2}\left( \begin{array}{cc}
a_0^{(1)}:\,a_1^{(1)},a_1^{(2)}; \ldots;a_r^{(1)},a_r^{(2)} & \\
&; x_1, \ldots, x_r\\
b_0^{(1)}:\,b_1^{(1)}; \ldots;b_r^{(1)}& \\
\end{array}\right) &=&\nonumber\\
=  \sum_{j_{1}, \ldots, j_r=0 }^{\infty} \frac{\left(a_0^{(1)}\right)_{j_1+\ldots+j_r}}{\left(b_0^{(1)}\right)_{j_1+\ldots+j_r}} \frac{\left(a_1^{(1)}\right)_{j_1}\left(a_1^{(2)}\right)_{j_1}\cdots\left(a_r^{(1)}\right)_{j_r}\left(a_r^{(2)}\right)_{j_r}}{\left(b_1^{(1)}\right)_{j_1}\left(b_r^{(1)}\right)_{j_r}}\frac{x_1^{j_1}x_2^{j_2}\cdots x_r^{j_r}}{j_1!j_2!\cdots j_r!}, & & \nonumber\\
\end{eqnarray}

Let us highlight that Eqs. (\ref{HSARE5})-(\ref{daoust}) give the angular contribution to the Rényi entropies not only for the hydrogenic systems, but also for any non-relativistic and spherically-symmetric multidimensional quantum system. Note from Eq. (\ref{HSARE5}) that the first term depends only on the spatial dimensionality and the second term on the first and last hyperquantum angular numbers, the spatial dimensionality and the entropic parameter $q$. Finally, the last term represents the sum of the different contributions of the internal angular degrees of freedom $\{ \mu_{j}\}_{j=2}^{D-2}$, which depends on the corresponding hyperquantum numbers $\mu_j$ and $\mu_{j+1}$.  
\noindent

\subsection{Total R\'enyi entropy in position space}

Finally, from Eqs. (\ref {eq:renyihyd1}), (\ref{HSRRE2}) and (\ref{HSARE5}) one has that the total Rényi entropy of the $D$-dimensional hydrogenic system in position space is given by
\begin{eqnarray}\label{Rqrho}
R_q[\rho_{n,l,\{\mu\}}]&=& D\ln \left(\frac{\pi^\frac12\eta}{2Z}\right)+\frac{q}{1-q}\ln\left(\frac{(\eta-L)_{2L+1}}{2\eta}  \right) 
\nonumber \\
&&+ \frac{1}{1-q}\ln\mathcal F_q(D,\eta,L)\, \mathcal A_q(D,L)
+\frac{1}{1-q}\ln\left[\frac{\Gamma(l+\frac D2)^q}{\Gamma\left(ql+\frac D2\right)}\frac{\Gamma\left(qm+1\right)}{\Gamma(m+1)^q}\right]\nonumber
\\
\nonumber \\
 & &+\frac{1}{1-q} \sum_{j=1}^{D-2} \ln \left[\mathcal B_q(D,\mu_j,\mu_{j+1})\,\,\mathcal G_q\left(D,\mu_j,\mu_{j+1}\right)\right]+\ln2
\end{eqnarray}
in terms of the hyperquantum numbers, the nuclear charge and the space dimensionality.

\subsection{Radial and total Rényi entropy in momentum space}

Operating in momentum space in a similar way as done for the position space in subsection 3.1, one has from Eqs. (\ref{eq:renentrop}), (\ref{eq:entropmom}) and (\ref{eq:momdens}) that the momentum radial Rényi entropy is given by 
\begin{eqnarray}
R_q[\gamma_{n,l}]&=&\frac1{1-q}\ln \left(\frac{Z^D}{\eta^D}\frac{K_{n,l}^{2q}}{2^{q(L
+2)}}\right)
\nonumber\\
&& +\frac1{1-q}\ln\int_{-1}^1 (1-y)^{lq+\frac D2-1} (1+y)^{ D(q-\frac12)+ q(l+1)-1} \mathcal C_{n-l-1}^{(L+1)}(y)^{2q}\,dy\nonumber
\end{eqnarray}
Again the use of the relation (\ref{relgegjac}) and  the Srivastava-Niukkanen-based linearization formula (\ref{linjac})  of the Jacobi polynomials has led us to find the following expression of the radial part of the R\'enyi entropy in momentum space:
\begin{eqnarray}
R_q[\gamma_{n,l}]&=& D\ln\frac{Z}\eta+\frac q{1-q}\ln \left[2\eta\,( \eta-L)_{2L+1}\right]
\\\nonumber
&+&\frac1{1-q}\ln \overline{\mathcal F}_q(D,\eta,L)+\frac1{1-q}\ln\overline{\mathcal A}_q(D,L)
\label{HSRREMS}
\end{eqnarray}
where 
\begin{eqnarray}
\overline{\mathcal F}_q(D,\eta,L)&\equiv& F_{1:1;\ldots;1}^{1:2;\ldots;2}\left( \begin{array}{cc}
a: b,c;\ldots;  b,c \\
&; 1, \ldots, 1\\
d: e; \ldots; e
\end{array}\right)
\nonumber\\
&&\hspace{-2cm}=\sum_{i_{1},\ldots,i_{2q}=0}^{n-l-1} \frac{(a)_{i_{1}+\ldots i_{2q} } }{(d)_{i_{1}+\ldots + i_{2q} } }\frac{ (b)_{i_{1} }(c)_{i_{1}} \cdots (b)_{i_{2q} }(c)_{i_{2q}}  }{(e)_{i_{1}}\cdots (e )_{i_{2q}} i_{1}!\cdots i_{2q}! }
\end{eqnarray}
with $a=(L+\frac32)q+\frac D2 (1-q)$, $b=-(\eta-L-1)$, $c=\eta+L+1$, $d=q(2L+4)$, $e=L+\frac32$ and 
\begin{equation}
\overline{\mathcal A}_q(D,L)\equiv2^{2q-1}\,\frac{\Gamma\left(\frac D2+ql\right)\,\Gamma\left(-\frac D2+q(D+l+1)\right)}{\Gamma\left(\frac D2+l\right)^{2q}\,\Gamma\left(q(D+2l+1)\right)}
\end{equation}
Note that, when $l=n-1$ the function $\mathcal F_q(D,\eta,L)=1$.

Finally, since the angular part of the momentum R\'enyi entropy is  the same as in position space, one obtains from Eqs. (\ref{HSRREMS})  and (\ref{HSARE5}) that the total Rényi entropy in momentum space    $R_q[\gamma_{\eta,L,\{\mu_j\}}] = R_q[\gamma_{n,l}] + R_{q}[\mathcal{Y}_{l,\{\mu \}}]$ has the following expression
\begin{eqnarray} \label{Rqgamma}
\hspace{-1cm}R_q[\gamma_{n,l,\{\mu\}}]&=& D\ln\left(\frac{\pi^\frac12Z}\eta\right)+\frac q{1-q}\ln \left[2\eta\,( \eta-L)_{2L+1}\right]
\nonumber
\\
 &&\hspace{-2.1cm} 
 +\frac{1}{1-q}\ln\left[\overline{\mathcal F}_q(D,\eta,L)\overline{\mathcal A}_q(D,L)\frac{\Gamma(l+\frac D2)^q}{\Gamma\left(ql+\frac D2\right)}\frac{\Gamma\left(qm+1\right)}{\Gamma(m+1)^q}\right]\hspace{-0.15cm}
\nonumber
\\
& & \hspace{-2.5cm}+\frac{1}{1-q} \sum_{j=1}^{D-2} \ln \left[\mathcal B_q(D,\mu_j,\mu_{j+1})\,\mathcal G_q\left(D,\mu_j,\mu_{j+1}\right)\right]+\ln 2
	\end{eqnarray}
in terms of the hyperquantum numbers, the nuclear charge and the space dimensionality.
\subsection{R\'enyi entropies for the quasi-spherical n\textit{s} states}

To illustrate the applicability of the previous position and momentum R\'enyi entropies, we calculate them for a relevant class of specific states of the $D$-dimensional hydrogenic system which include the ground state: the quasi-spherical n\textit{s} states, which are characterized by the angular hyperquantum numbers $\mu_1=\mu_2\ldots=\mu_{D-1}=l$. First, since $l=n-1$, the Lauricella function of Eq. (\ref{HSRRE2}) is equal to unity. Then, we find the values \begin{equation}\nonumber
\label{HSRREns}
R_{q}[\rho_{n,n-1}] = D\ln \frac{\eta}{2Z}-	\frac{q}{1-q}\ln\left[\Gamma\left(2\eta+1\right)  \right]+\frac{1}{1-q}\ln\left(\frac{\Gamma\left(D+2nq-2q\right)}{q^{D+2nq-2q}} \right)
\end{equation}
for the radial R\'enyi entropy of the n\textit{s} states in position space, and 
\begin{eqnarray}\nonumber
\label{HSRREgs}
R_{q}[\rho_{1,0}] &=& \Gamma (D)+D\ln\left[ \frac{D-1}{4Z\,q^{\frac1{1-q}}}\right]
\end{eqnarray}
for the corresponding one of the ground state (n=1). In addition, we have found the values 
\begin{eqnarray}
\label{HSAREns}\nonumber
\hspace{-1cm}R_{q}[\mathcal{Y}_{l,\{\mu \}}]  &=&\ln(2\pi^{\frac D2})+\frac{1}{1-q}\ln \left[\frac{\Gamma(l+\frac D2)^q}{\Gamma(l+1)^q}\frac{\Gamma\left(ql+1\right)}{\Gamma\left(ql+\frac D2\right)}\right]
\end{eqnarray}
and  
\begin{eqnarray}
\label{HSAREgs}\nonumber
\hspace{-1cm}R_{q}[\mathcal{Y}_{0,\{0 \}}]  &=&\ln\left[\frac{2\,\pi^{\frac D2}}{\Gamma\left(\frac D2\right)}\right]
\end{eqnarray}
for the angular R\'enyi entropy of the n\textit{s} states and the ground state, respectively. 
%
Similar operations in the momentum space have allowed us to have the values
\begin{eqnarray}
\label{HSRREMSns}\nonumber
R_q[\gamma_{n,n-1}]&=& D\ln\frac{Z}\eta+\frac q{1-q}\ln\left[ 4\,\Gamma(2\eta+1)\right]
\\
&&\hspace{-1cm}+\frac1{1-q}\ln\left[\frac{\Gamma\left(\frac D2+qn-q\right)\,\Gamma\left(-\frac D2+q(D+n)\right)}{2\Gamma\left(n+\frac{D}2-1\right)^{2q}\Gamma\left(q(D+2n-1)\right)}\right]
\nonumber
\end{eqnarray}
and
\begin{eqnarray}
\label{HSRREMSgs}\nonumber
R_q[\gamma_{1,0}] &=& D\ln\left[\frac{2Z}{D-1} \right]+\frac q{1-q}\ln\left[ 4\,\Gamma(D)\right]
\\
&+&\frac1{1-q}\ln\left[\frac{\Gamma\left(\frac D2\right)^{1-2q}\,\Gamma\left(D(q-\frac 12)+q\right)}{2\Gamma\left(Dq+q\right)}\right]
\nonumber
\end{eqnarray}
for for the radial R\'enyi entropy of the n\textit{s} states and the ground state in momentum space, respectively.\\

Finally, we gather in Tables 1 and 2 the exact values of the position and momentum R\'enyi entropies $R_2[\rho_{n,l,m}]$ and $R_2[\gamma_{n,l,m}]$, respectively, of various quasi-circular circular states of the three-dimensional hydrogen atom. Therein, we observe that this quantity increases (decreases) when the main quantum number $n$ is increasing (decreasing) in position space, and the opposite behavior is observed in momentum space. Moreover, it decreases (increases) when the orbital quantum number $l$ is increasing (decreasing) in position space, and the opposite behavior is observed in momentum space. And, the associated joint position-momentum entropy sum increases in a systematic way when $n$ or $l$ are increasing.

\begin{table}[h!]
\begin{center}
\setlength{\tabcolsep}{7.0pt}
\renewcommand{\arraystretch}{1.5}
\begin{tabular}{|c|c|c|c|}
 \hline
 $R_2[\rho_{n,l,m}]$ &  $n=1$ & $n=2$ & $n=3$
 \\
 \hline
 $l=0, m=0$ & $\ln(8\pi)$ & $\ln\left(\frac{2048\,\pi}5\right)$  & $\ln\left(\frac{20736\,\pi}5\right)$ 
 \\
 \hline
 $l=1,m=0$ &-	& $\ln\left(\frac{2048\,\pi}9\right)$ &$\ln\left(\frac{27648\,\pi}{11}\right)$
 \\
 \hline
 $l=1,m=1$ &-	& $\ln\left(\frac{1024\,\pi}3\right)$&$\ln\left(\frac{41472\,\pi}{11}\right)$ 
 \\
 \hline
$l=2,m=0$ &- &- &$\ln\left(\frac{9216\,\pi}{5}\right)$
\\
\hline
$l=2,m=1$ &- &- &$\ln\left(\frac{13824\,\pi}{5}\right)$
\\
\hline
$l=2,m=2$ &- &- &$\ln\left(\frac{13824\,\pi}{5}\right)$
\\
\hline
\end{tabular}
\end{center}
\caption{Exact values of the total position Rényi entropy $R_2[\rho_{n,l,m}]$ for various quasi-circular states of the three-dimensional hydrogen atom.}
\label{table1}
\end{table}
\begin{table}[h!]
\begin{center}
\setlength{\tabcolsep}{7.0pt}
\renewcommand{\arraystretch}{1.5}
\begin{tabular}{|c|c|c|c|}
 \hline
 $R_2[\gamma_{n,l,m}]$ &  $n=1$ & $n=2$ & $n=3$
 \\
 \hline
 $l=0, m=0$ & $\ln(\frac{16\pi^2}{33})$ & $\ln\left(\frac{2\pi^2}{151}\right)$  & $\ln\left(\frac{16\pi^2}{7533}\right)$ 
 \\
 \hline
 $l=1,m=0$ &-	& $\ln\left(\frac{2\pi^2}{39}\right)$ &$\ln\left(\frac{160\,\pi^2}{36207}\right)$
 \\
 \hline
 $l=1,m=1$ &-	& $\ln\left(\frac{\pi^2}{13}\right)$&$\ln\left(\frac{80\,\pi^2}{12069}\right)$
 \\
 \hline
$l=2,m=0$ &- &- &$\ln\left(\frac{1120\,\pi^2}{78489}\right)$
\\
\hline
$l=2,m=1$ &- &- &$\ln\left(\frac{560\,\pi^2}{26163}\right)$
\\
\hline
$l=2,m=2$ &- &- &$\ln\left(\frac{560\,\pi^2}{26163}\right)$
\\
\hline
\end{tabular}
\end{center}
\caption{Exact values of the total momentum Rényi entropy $R_2[\gamma_{n,l,m}]$ for various quasi-circular states of the three-dimensional hydrogen atom.}
\label{tablemom}
\end{table}

\section{Position-momentum Rényi-entropy sum}

Here we give the joint position-momentum R\'enyi uncertainty sum for all the discrete stationary states of the $D$-dimensional hydrogenic system from Eqs. (\ref {Rqrho}) and (\ref{Rqgamma}). We obtain
\begin{eqnarray}\label{RenyiSum}
R_q[\rho_{n,l,\{\mu\}}]+R_p[\gamma_{n,l,\{\mu\}}]&=& D\ln\left(\frac{\pi}2\right)+\frac {2q}{q-1}\ln \left[2\eta\right]
\nonumber
\\
 &&\hspace{-5cm}
  +\ln\left[\mathcal F_q(D,\eta,L)^{\frac{1}{1-q}}\overline{\mathcal F}_p(D,\eta,L)^{\frac{1}{1-p}}\,\mathcal A_q(D,L)^{\frac{1}{1-q}}\overline{\mathcal A}_p(D,L)^{\frac{1}{1-p}}\right]\hspace{-0.15cm}
\nonumber
\\
 &&\hspace{-5cm} 
 +\ln\left[\left(\frac{\Gamma\left(qm+1\right)}{\Gamma\left(ql+\frac D2\right)}\right)^{\frac{1}{1-q}}\left(\frac{\Gamma\left(pm+1\right)}{\Gamma\left(pl+\frac D2\right)}\right)^{\frac1{1-p}}\right]
\nonumber
\\
& & \hspace{-5cm}+ \sum_{j=1}^{D-2} \ln \left[(q\mu_{j+1}+\alpha_j+1)_{q(\mu_{j+1}-\mu_j)}^{\frac1{1-q}}\,(p\mu_{j+1}+\alpha_j+1)_{p(\mu_{j+1}-\mu_j)}^{\frac1{1-p}}\right]
\nonumber\\
& & \hspace{-5cm}+ \sum_{j=1}^{D-2} \ln \left[\left(\mathcal G_q(D,\mu_j,\mu_{j+1})\right)^\frac1{1-q}\,\left(\mathcal G_p\left(D,\mu_j,\mu_{j+1}\right)\right)^\frac1{1-p}\right]+\ln 4
\end{eqnarray}
with $ \frac1p+\frac1q=2$\footnote{In fact, this is only valid provided the functions $\mathcal F_q,\overline{\mathcal F_q}$ and $\mathcal G_q$ exist for any $p,q\in \mathbb R$.}.
When the spatial dimension is $D=3$ this expression boils down to
\begin{eqnarray}
R_q[\rho_{n,l,m}]+R_p[\gamma_{\eta,L,\{\mu_j\}}]&=& 3\ln\left(\frac{\pi}2\right)+\frac {2q}{q-1}\ln \left[2n\right]
\nonumber
\\
 &&\hspace{-5cm}
  +\ln\left[\mathcal F_q(3,n,l)^{\frac{1}{1-q}}\overline{\mathcal F}_p(3,n,l)^{\frac{1}{1-p}}\,\mathcal A_q(3,l)^{\frac{1}{1-q}}\overline{\mathcal A}_p(3,l)^{\frac{1}{1-p}}\right]\hspace{-0.15cm}
\nonumber
\\
& & \hspace{-5cm}+\ln \left[\left(\mathcal G_q(3,l,m)\right)^\frac1{1-q}\,\left(\mathcal G_p\left(3,l,m\right)\right)^\frac1{1-p}\right]
\nonumber
\\
&&\hspace{-5cm} 
 +\ln\left[4\left(\frac{\Gamma\left(qm+1\right)}{\Gamma\left(qm+\frac 32\right)}\right)^{\frac{1}{1-q}}\left(\frac{\Gamma\left(pm+1\right)}{\Gamma\left(pm+\frac 32\right)}\right)^{\frac1{1-p}}\right]
\end{eqnarray}
Finally and most interesting, the expression (\ref{RenyiSum}) in the limit $D\to\infty$ becomes
\begin{equation}
R_q[\rho_{\eta,L,\{\mu_j\}}]+R_p[\gamma_{\eta,L,\{\mu_j\}}]\sim D\ln\left(2\pi (2q)^{\frac{1}{2-2q}}(2p)^{\frac{1}{2-2p}}\right),
\end{equation}
which is the saturation value of the R\'enyi-entropy-based uncertainty relation found independently by Bialynicki-Birula \cite{bialynicki2} and Zozor-Portesi-Vignat \cite{vignat,portesi}.
 \begin{equation}
 \label{sum}
R_q[\rho]+R_p[\gamma]\ge D\ln\left(2\pi (2q)^{\frac{1}{2-2q}}(2p)^{\frac{1}{2-2p}}\right),\quad\frac1p+\frac1q=2.
 \end{equation}
This fact is not only a partial checking of our results but also it is in accordance with similar findings  obtained in a very different way.

\newpage

\section{Conclusions}

In this work we have explicitly calculated the total position $R_q [\rho_{n,l,\{\mu\}}]$ and momentum $R_q [\gamma_{n,l,\{\mu\}}]$ R\'enyi entropies (with integer $q$ greater than 1) for all the 
quantum-mechanically allowed hydrogenic states  in terms of the Rényi parameter $q$, the spatial dimension $D$, the nuclear charge $Z$ as well as
the hyperquantum numbers, $(n,l,\{\mu \})$, which characterize the corresponding wavefunction of the states. We have learnt for instance that the R\'enyi entropies behaves as $D\ln \left(\frac{\pi^\frac12\eta}{2Z}\right)$ and $D\ln\frac{Z}\eta$ in the position and momentum spaces, respectively, so that the joint R\'enyi uncertainty measure linearly grows with the space dimensionality and it does not depend on the nuclear charge. 
Moreover, we have analytically shown the dependence of the R\'enyi entropies on the hyperquantum numbers, such as we have illustrated for the three-dimensional hydrogenic quasi-circular states; in the latter case we observe, in particular, that the joint position-momentum entropy sum is an increasing function of the quantum numbers.\\

We have used a recent methodology which allows to determine the involved integral functionals by taking into account the linearization formula and orthogonality conditions of the Laguerre and Jacobi polynomials; the latter ones are closely connected to the Gegenbauer polynomials which control the angular part of the wavefunctions in both conjugated spaces as well as the radial wavefunction in momentum space. The final expressions for the Rényi entropies in position and momentum spaces are expressed in a compact way by use of a multivariate hypergeometric function of Lauricella and Srivastava-Daoust types evaluated at $1/q$ and unity, respectively; indeed, note that all sums to be evaluated are finite. Moreover, as a byproduct, we have been able to obtain the exact value of the angular contribution for the Rényi entropies of any non-relativistic and spherically-symmetric multidimensional quantum system.
Finally, it remains as an open problem the extension of this result to the limiting case $q \to 1$, which corresponds to the Shannon entropy, and the R\'enyi entropies for any real value of the parameter $q$. The latter requires a completely different approach, still unknown to the best of our knowledge.  

\section*{Acknowledgments}
This work has been partially supported by the Projects FQM-7276 and FQM-207 of the Junta de Andaluc\'ia and the MINECO-FEDER grants FIS2014- 54497P, FIS2014-59311P and FIS2017-89349-P. 
\end{document}